\documentclass[aps,twocolumn,preprintnumbers,amsmath,amssymb,nofootinbib,superscriptaddress,notitlepage]{revtex4}
\usepackage{graphicx,color,dcolumn,booktabs,bm}
\usepackage{longtable,lscape}
\usepackage{txfonts}
\usepackage{overpic}
\usepackage{amssymb}
\usepackage{indentfirst}
\usepackage{feynmf}
\usepackage{slashed}
\usepackage{cases}
\usepackage{multirow}
\usepackage{appendix}
\usepackage[figuresright]{rotating}
\usepackage[colorlinks,
citecolor=blue,
anchorcolor=red,
menucolor=red,
linkcolor=red,
filecolor=red,
runcolor=red,
urlcolor=blue,
frenchlinks=red]{hyperref}
\usepackage{epstopdf}
\usepackage{bm}
\usepackage{tabularx}
\usepackage{threeparttable}
\usepackage{bbding}
\usepackage{array}
\usepackage[section]{placeins}
\usepackage{makecell}
\usepackage{comment}
\makeatletter

\newcommand{\Rmnum}[1]{\expandafter\@slowromancap\romannumeral #1@}
\makeatother

\usepackage{soul}
\begin{document}
	\title{How higher charmonia shape the puzzling data of the $e^+e^-\to \eta J/\psi$ cross section}
	\author{Tian-Cai Peng}
	\email{pengtc20@lzu.edu.cn}
	\author{Zi-Yue Bai}
	\email{baizy15@lzu.edu.cn}
	\affiliation{School of Physical Science and Technology, Lanzhou University, Lanzhou 730000, China}
	\affiliation{Lanzhou Center for Theoretical Physics, Key Laboratory of Theoretical Physics of Gansu Province, Lanzhou University, Lanzhou 730000, China}
	\affiliation{Key Laboratory of Quantum Theory and Applications of MoE, Lanzhou University, Lanzhou 730000, China}
	\affiliation{Research Center for Hadron and CSR Physics, Lanzhou University and Institute of Modern Physics of CAS, Lanzhou 730000, China}

	\author{Jun-Zhang~Wang}
	\email{wangjzh2022@pku.edu.cn}
	\affiliation{School of Physical Science and Technology, Lanzhou University, Lanzhou 730000, China}
	\affiliation{School of Physics and Center of High Energy Physics, Peking University, Beijing 100871, China}
	
	\author{Xiang~Liu}
	\email{xiangliu@lzu.edu.cn}
	\affiliation{School of Physical Science and Technology, Lanzhou University, Lanzhou 730000, China}
	\affiliation{Lanzhou Center for Theoretical Physics, Key Laboratory of Theoretical Physics of Gansu Province, Lanzhou University, Lanzhou 730000, China}
	\affiliation{Key Laboratory of Quantum Theory and Applications of MoE, Lanzhou University, Lanzhou 730000, China}
	\affiliation{Research Center for Hadron and CSR Physics, Lanzhou University and Institute of Modern Physics of CAS, Lanzhou 730000, China}
	\affiliation{MoE Frontiers Science Center for Rare Isotopes, Lanzhou University, Lanzhou 730000, China}

	\date{\today}
	
	\begin{abstract}
		Recently, the BESIII collaboration performed a precise measurement of the $e^+e^-\to \eta J/\psi$ cross section.
		It is puzzling that the resonance parameters of the reported $Y(4230)$ show a substantial divergence from the previously measured results in both the open-charmed and hidden-charmed decay channels, and the line shape asymmetry of the data approaching 4.2 GeV also suggests that it might be difficult to characterize the details of the structure around 4.2 GeV by a single resonance. This has motivated our great curiosity about how the charmonium states are distributed in the measured energy range and how they shape the puzzling data of the $e^+e^-\to \eta J/\psi$ cross section. In this work, we use five theoretically constructed charmonia in the range of $4.0\rm{-}4.5$ $\text{GeV}$, i.e., $\psi(4040)$, $\psi(4160)$, $\psi(4220)$, $\psi(4380)$, and $\psi(4415)$, to apply a combined fit to the data, in which their calculated decay ratios into $\eta J/\psi$ via hadronic loop mechanism  are taken as input. The fit results can reproduce the measured cross section data well, especially for the subtle line shape around 4.2 GeV, showing that the structure around 4.2 GeV is possible from the contribution of both $\psi(4160)$ and $\psi(4220)$.

	\end{abstract}
	
	\maketitle
	
	\section{INTRODUCTION}
	With the continuous accumulation of data, an increasing number of charmoniumlike $XYZ$ states have been documented in various experimental endeavors, such as Belle, BESIII, LHCb, and CMS. This surge in reported states has invigorated the field of hadron spectroscopy, giving it a remarkable level of activity and importance (see review articles \cite{Liu:2013waa,Chen:2016qju,Esposito:2016noz,Chen:2016spr,Guo:2017jvc,Olsen:2017bmm,Liu:2019zoy,Brambilla:2019esw,Chen:2022asf} for more details). This research landscape has led to significant advances in our construction of hadron spectroscopy. In particular, it has advanced our understanding of the intricate nonperturbative dynamics associated with the strong interactions, which is a central frontier of contemporary physics research.
	
	Recently, the BESIII collaboration measured the $e^+e^- \to \eta J/\psi$ cross section over a center-of-mass energy range of $\sqrt{s} = 3.808\rm{-}4.951$ GeV \cite{BESIII:2023tll}. This measurement was made by analyzing a data sample with an integrated luminosity of 22.42 $\rm{fb}^{-1}$~\cite{BESIII:2023tll}. To depict this data, BESIII introduced three resonant structures, i.e., the charmonium state $\psi(4040)$ and two charmoniumlike states, namely $Y(4230)$ and $Y(4360)$. The resonance parameters of these introduced states were determined. For the $Y(4230)$ state, the resonance mass was found to be $m_{Y(4230)} = 4219.7 \pm 2.5 \pm 4.5$ MeV, and the corresponding width was measured to be $\Gamma_{Y(4230)} = 80.7 \pm 4.4 \pm 1.4$ MeV. Similarly, the $Y(4360)$ state showed a mass of $m_{Y(4360)} = 4386 \pm 13 \pm 17$ MeV, accompanied by a width of $\Gamma_{Y(4360)} = 177 \pm 32 \pm 13$ MeV~\cite{BESIII:2023tll}. It is worth noting that the observed width of the $Y(4230)$ state, as reported, exceeds those found in other hidden-charm and open-charm decay channels. These include processes such as $e^+e^- \to \pi^+\pi^-J/\psi$~\cite{BaBar:2005hhc,BESIII:2016bnd}, $\pi^0\pi^0J/\psi$~\cite{BESIII:2020oph}, $\psi(2S) \pi^+\pi^-$~\cite{BaBar:2006ait},  $\pi^+\pi^-\psi(3686)$~\cite{BESIII:2017tqk}, $h_c \pi^+\pi^-$~\cite{BESIII:2016adj}, $\chi_{c0} \omega$~\cite{BESIII:2019gjc}, $K^+K^-J/\psi$\cite{BESIII:2022joj}, and $D^0D^{*-}\pi^+$\cite{BESIII:2018iea}.
	This difference in width hints at some distinct nature of this reported $Y(4230)$ structure in $e^+e^- \to \eta J/\psi$ \cite{BESIII:2023tll}.

	As collected by the Particle Data Group \cite{ParticleDataGroup:2022pth}, the family of charmonium encompasses two well-established entities, namely  $\psi(4160)$ and $\psi(4415)$, in addition to $\psi(4040)$. However, upon meticulous examination of the experimental data reported by BESIII, a conspicuous absence is observed within the cross section distribution of $e^+e^- \to\eta J/\psi $, pertaining specifically to the aforementioned charmonium states, $\psi(4160)$ and $\psi(4415)$. Remarkably, the established charmonium $\psi(4160)$ possesses sizable decay fractions into the $\eta J/\psi $ channel according to the theoretical calculation \cite{Chen:2012nva}. Moreover, if carefully checking the BESIII's data, we may find an asymmetric line shape, associated with the resonance structure denoted as $Y(4230)$. Confronted by this puzzling phenomenon existing in the cross section data of $e^+e^-\to\eta J/\psi $ \cite{BESIII:2023tll}, it is necessary to come up with a convincing solution to this problem.
	We should mention the distinct characterized energy level structure of higher charmonia within the framework of the unquenched picture,  a contribution credited to the diligent efforts of the Lanzhou group \cite{Wang:2019mhs,Wang:2022jxj,Wang:2023zxj}. In the newly constructed $J/\psi$ family, there are six vector charmonium states in the energy range of $4\rm{-}4.5$ GeV, i.e., $\psi(4040)$, $\psi(4160)$, $\psi(4220)$, $\psi(4380)$, $\psi(4415)$, and $\psi(4500)$ \cite{Wang:2019mhs}. In this work, we reveal how these 
	higher charmonia shape the puzzling data of the $e^+e^-\to \eta J/\psi$ cross section.
	
	This paper is organized as follows. 
	After the Introduction, we briefly review the status of these higher charmonia in the range of $4.0\rm{-}4.5$ $\text{GeV}$, which are used to solve the puzzles appearing in the cross section data of the $e^+e^-\to J/\psi\eta$ in Sec. \ref{SecII}. We then illustrate the calculation details of these charmonium decays into $\eta J/\psi$ via the hadronic loop mechanism in Sec. \ref{SubA}. Based on these results, we perform a combined fit to show how these 
	higher charmonia shape the puzzling data of the $e^+e^-\to \eta J/\psi$ cross section in Sec. \ref{SubB}. The paper ends with a short summary in Sec. \ref{SecIII}.
	
	\section{An analysis to the cross section of  $e^+e^-\to \eta J/\psi$}\label{SecII}
	
	The continued accumulation of the experimental data in the range of $4.0\rm{-}4.5$ $\text{GeV}$ has deepened our understanding of the spectrum of higher vector charmonium above 4.0 GeV. Recently, the BESIII collaboration performed a precise measurement of the $e^+e^-\to \eta J/\psi$ cross section over a center-of-mass energy range of $\sqrt{s} = 3.808\rm{-}4.951$ GeV, and obtained the charmonium state $\psi(4040)$ and the other two charmoniumlike states named $Y(4230)$ and $Y(4360)$ from a Breit-Wigner fit, but the extracted resonance parameters of the observed $Y(4230)$ state differ from other measurements of the open-charmed and hidden-charmed decay channels \cite{BaBar:2005hhc,BESIII:2016bnd,BESIII:2020oph,BaBar:2006ait,BESIII:2017tqk,BESIII:2016adj,BESIII:2019gjc,BESIII:2022joj,BESIII:2018iea}, so we can conclude that the $\psi(4230)$ resonance parameters can be influenced by the contribution of the nearby $\psi(4160)$ state.
	The line shape of the observed $Y(4230)$ looks asymmetric and irregular, and could not be formed by the contribution of just one resonance. It seems that the structure around 4.2 GeV from the measured cross section data of $e^+e^- \to \eta J/\psi$ may have some substructures, which is also endorsed in a recent theoretical study conducting a global coupled-channel analysis for the BESIII's data \cite{Nakamura:2023obk}. 
	
	
	In order to fully understand these puzzling data, the theoretical inputs of the vector charmonia located in the range of $4.0\rm{-}4.5$ $\text{GeV}$ are crucial.
	Besides the well constructed charmonia, i.e., $\psi(4040)$, $\psi(4160)$ and $\psi(4415)$ in the range of $4.0\rm{-}4.5$ $\text{GeV}$, the Lanzhou group indicated that there should exist a narrow charmonium $\psi(4220)$ corresponding to $Y(4220)$ reported in the processes of $e^+e^-\to J/\psi \pi^+\pi^-$, $e^+e^-\to h_c\pi^+\pi^-$, $e^+e^-\to\chi_{c0}\omega$ and $e^+e^-\to \psi(3686)\pi^+\pi^-$ \cite{He:2014xna,Chen:2014sra,Chen:2015bma,Chen:2017uof}. 
	In Ref. \cite{Wang:2019mhs}, the Lanzhou group used an unquenched potential model with $\psi(4220)$ as the scaling point and predicted another two more charmonium states named $\psi(4380)$ and $\psi(4500)$.  The information of these charmonia is summarized in Table \ref{tab: parameter}. It is worth mentioning that the BESIII experiment recently indeed found a new structure around 4.5 GeV in the measurement of $e^+e^-\to K^+K^-J/\psi$ \cite{BESIII:2022joj}, which just can relate to our predicted charmonium $\psi(4500)$.  With these six vector charmonia, theoretically established in the range of $4.0\rm{-}4.5$, i.e., $\psi(4040)$, $\psi(4160)$, $\psi(4220)$, $\psi(4380)$, $\psi(4415)$ and $\psi(4500)$, they have shown that the experimental cross section measurements of $e^+e^-\to\psi(2S)\pi^+\pi^-$ \cite{Wang:2019mhs}, $e^+e^-\to K^+K^-J/\psi$ \cite{Wang:2022jxj} and $e^+e^-\to\pi^+D^0D^{*-}$ \cite{Wang:2023zxj} can be explained under a unified charmonium spectroscopy. In this work, we aim to show that the newly measured $e^+e^- \to \eta J/\psi$ cross section data can also be understood by the contributions of the same energy level structures, and how they shape the puzzling $e^+e^-\to \eta J/\psi$ cross section data.
	
	The contribution of genuine intermediate charmonium to $e^+e^- \to \eta J/\psi$ can be described by a phase space corrected Breit-Wigner function, i.e.,
	\begin{align}\label{a}
		\mathcal{M}_{\psi}(s) =&\frac{\sqrt{12\pi\Gamma_\psi^{e^+e^-}\mathcal{B}(\psi\to \eta J/\psi)\Gamma_\psi}}{s-m_\psi^2+im_\psi\Gamma_\psi}  \\\nonumber
		&\times\sqrt{\frac{\Phi_{2\to2}(s)}{\Phi_{2\to2}(m_{\psi}^2)}},
	\end{align}
	where $m_{\psi}$, $\Gamma_{\psi}$, and $\Gamma_\psi^{e^+e^-}$ are the mass, total width and the dielectron width of the intermediate charmonium, respectively. $\Phi_{2\to 2}$ is the phase space and $s$ donates the center-of-mass energy. 
	The only remaining unknown term $\mathcal{BR}(\psi\to \eta J/\psi)$ is the branching ratio of the associated charmonium decay into $\eta J/\psi$, and we will present the details of the calculation for the $\psi\to \eta J/\psi$ decay next.
	
	\begin{table}[ht]
		\caption{The theoretical mixing angles, assignments, masses, total widths, and dielectron widths of higher charmonia in the range of $4.0\rm{-}4.5$ $\text{GeV}$, which are obtained from the theoretical predictions \cite{Wang:2019mhs,Wang:2022jxj,Wang:2023zxj} and some experimental values which are close to the theoretical predictions.}
		\label{tab: parameter}
		\setlength{\tabcolsep}{1.5pt}
		\centering 
		\begin{tabular}{cccccc}
			\hline \hline
			States &~$\theta$  &Assignment  &Mass (MeV) &$\Gamma$ (MeV) &$\Gamma_\psi^{e^+e^-}$ (keV)\\ \hline
			$\psi(4040)$ &~$20^\circ$ &$\psi^{\prime}_{3S-2D}$		&$4039\pm1$ \cite{ParticleDataGroup:2022pth} &$80\pm10$ \cite{ParticleDataGroup:2022pth} &0.830 \cite{BES:2007zwq}\\
			$\psi(4160)$ &~$20^\circ$ &$\psi^{\prime\prime}_{3S-2D}$ &$4159\pm20$ \cite{DASP:1978dns} &$78\pm20$ \cite{DASP:1978dns} &0.480 \cite{BES:2007zwq}\\
			$\psi(4220)$ &~$34^\circ$ &$\psi^{\prime}_{4S-3D}$ 		&4222 &44 &0.290\\
			$\psi(4380)$ &~$34^\circ$ &$\psi^{\prime\prime}_{4S-3D}$ &4389 &80 &0.257\\
			$\psi(4415)$ &~$30^\circ$ &$\psi^{\prime}_{5S-4D}$ 		&4414 &33 &0.230\\
			$\psi(4500)$ &~$30^\circ$ &$\psi^{\prime\prime}_{5S-4D}$ &~4509&50 &0.113\\
			\hline \hline 	
		\end{tabular}
	\end{table}

	\subsection{Calculating the branching ratios of higher charmonium decays into $\eta J/\psi$}\label{SubA}

	In Ref. \cite{Wang:2019mhs}, an $S$-$D$ mixing scheme was proposed to construct the energy level structure of vector charmonia in the range of $4.0\rm{-}4.5$ $\text{GeV}$, as shown below,
	
	\begin{equation}\label{mix}
	\left(\begin{array}{c}
	|\psi^\prime\rangle \\
	|\psi^{\prime\prime}\rangle \end{array}\right)=
	\left(\begin{array}{cc}\cos\,\theta&\sin\,\theta\\
	-\sin\,\theta&\cos\,\theta\end{array}\right)
	\left(\begin{array}{c}|\psi(nS)\rangle\\
	|\psi((n-1)D)\rangle
	\end{array}\right),
	\end{equation}
	where $\theta$ denotes the mixing angle.
	
	In this scheme, $\psi(4220)$ was assigned to a $4S$-$3D$ mixing state while its partner $\psi(4380)$ was predicted. The well-established $\psi(4415)$ was assigned into a $5S$-$4D$ mixing state while its partner $\psi(4500)$ was predicted and was found to exist in the BESIII measurement of $e^+e^-\to K^+K^-J/\psi$ \cite{Wang:2022jxj}. The assignment of $\psi(4040)$ and $\psi(4160)$ with a $3S$-$2D$ mixing scheme has also been proposed in Ref. \cite{Wang:2022jxj} to match the experimental dielectron width of $\psi(4160)$ \cite{BES:2007zwq}. The mixing angles are listed in Table \ref{tab: parameter}. Next, we will utilize the hadronic loop mechanism to calculate the branching ratios of the hidden-charm decays of $\psi \to \eta J/\psi$ for the above charmonia.
	
\begin{figure}[htbp]\centering
		\includegraphics[width=86mm]{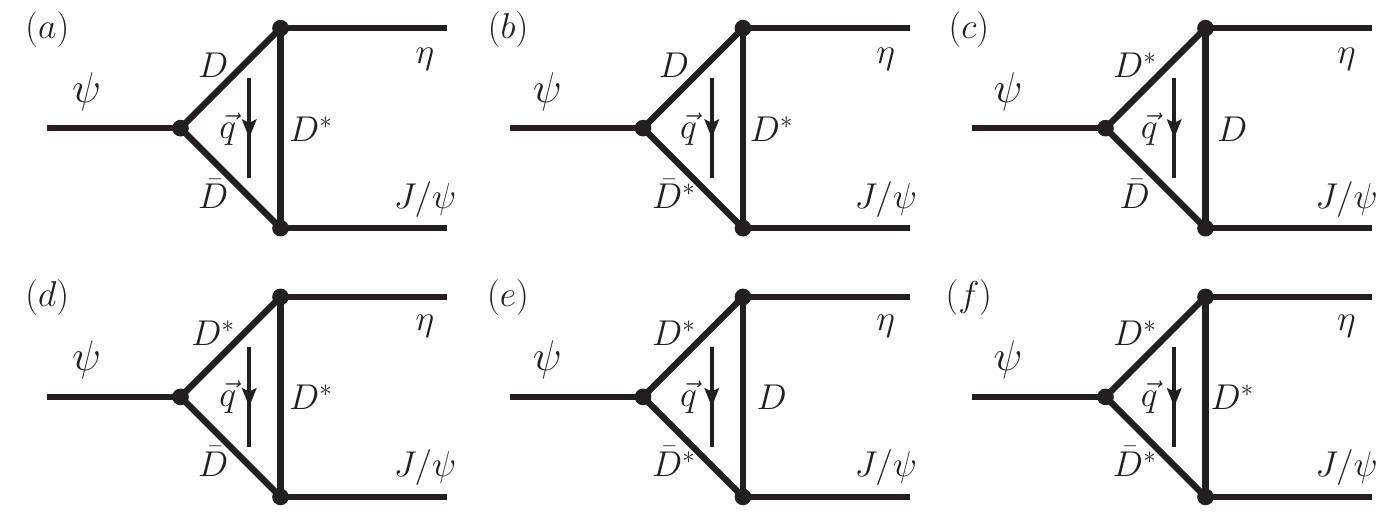}\\
		\caption{The schematic diagrams illustrate the hidden-charm decay decay of higher charmonium into $\eta J/\psi$ via the hadronic loop mechanism, and six sub-diagrams ((a)-(f)) represent different $D^{(*)}$ meson loop processes.}
		\label{fig: hadron loop}
	\end{figure}

	The hadronic loop mechanism has been widely used to study the hidden-flavor decays of heavy quarkonia above the open-flavor thresholds \cite{Cheng:2004ru,Liu:2006dq,Liu:2009dr,Meng:2007tk,Meng:2008dd,Meng:2008bq,Chen:2011qx,Chen:2011zv,Chen:2011pv,Chen:2011jp,Chen:2014ccr,Wang:2015xsa,Wang:2016qmz,Huang:2017kkg,Zhang:2018eeo,Huang:2018cco,Huang:2018pmk,Li:2021jjt,Bai:2022cfz,Li:2022leg,Bai:2023dhc,Li:2013zcr} and the calculated branching ratios are usually comparable to the experimental measurements. In the framework of the hadronic loop mechanism, the higher charmonium $\psi^\prime/\psi^{\prime\prime}$ within an $S$-$D$ mixture first decay into a pair of charmed mesons $D^{(*)}\bar D^{(*)}$, and reach the $\eta J/\psi$ final states by exchanging a $D$ or $D^*$ meson as shown in Fig. \ref{fig: hadron loop}. The general expression for the amplitude mediated by the charmed meson loop is
	
	\begin{align}
		{\cal M}=\int\frac{d^{4}q}{(2\pi)^{4}}\frac{{\cal V}_{1}{\cal V}_{2}{\cal V}_{3}}{{\cal P}_{1}{\cal P}_{2}{\cal P}_{E}}{\cal F}^{2}(q^{2},m_{E}^{2}), \label{eq3}
	\end{align}
	where ${\cal V}_{i}(i=1,2,3)$ are interaction vertices, and ${\cal P}_{i}(i=1,2,E)$ denote the corresponding propagators of intermediate charmed mesons. The form factor $\mathcal{F}(q^{2},m_{E}^{2})$ is introduced to compensate for the off-shell effect of the exchanged $D^{(*)}$ meson and to depict the structure effect of the interaction vertices. In our calculation, the monopole form factor is taken as
	
	\begin{align}\label{FFs}
		\mathcal{F}(q^{2},m_{E}^{2})=\frac{\Lambda^{2}-m_{E}^{2}}{\Lambda^{2}-q^{2}},
	\end{align}
	where $m_E$ and $q$ are the mass and four momentum of the exchanged intermediate meson, respectively. The cutoff $\Lambda$ can be parametrized as $\Lambda=m_{E}+\alpha\Lambda_{QCD}$, with $\Lambda_{QCD}=220$ MeV \cite{Liu:2006dq,Liu:2009dr,Li:2013zcr}, $\alpha$ is usually of order of 1 and depends on the specific processes \cite{Cheng:2004ru}.

	The effective Lagrangian approach is used to give the concrete expressions for the decay amplitudes defined in Eq.~(\ref{eq3}). The Lagrangians of the concrete interactions involved in Fig. \ref{fig: hadron loop} are listed below \cite{Li:2021jjt},
	
	\begin{align}\label{psi_S_wave to DD}
		\mathcal{L}_{\psi_{(S)}\mathcal{D}^{(*)}\mathcal{D}^{(*)}}=
		&ig_{\psi_{(S)}\mathcal{DD}}\psi^{\mu}(\partial_{\mu}\mathcal{D}^{\dagger}\mathcal{D}-\mathcal{D}^{\dagger}\partial_{\mu}\mathcal{D})  \\\nonumber
		&+g_{\psi_{(S)}\mathcal{DD}^{*}}\varepsilon_{\mu\nu\alpha\beta}\partial^{\mu}\psi^{\nu}(\mathcal{D}^{*\alpha\dagger}\overleftrightarrow{\partial}^\beta\mathcal{D}-\mathcal{D}^{\dagger}\overleftrightarrow{\partial}^\beta\mathcal{D}^{*\alpha}) \\\nonumber
		&+ig_{\psi_{(S)}{\mathcal D}^{*}{\mathcal D}^{*}}\psi^{\mu}(\partial_{\nu}{\mathcal D}_{\mu}^{*\dagger}{\mathcal D}^{*\nu}-{\mathcal D}^{*\nu\dagger}\partial_{\nu}{\mathcal D}_{\mu}^{*}\\\nonumber
		&+{\mathcal D}^{*\nu\dagger}\partial_{\mu}{\mathcal D}_{\nu}^{*}),\\
		\mathcal{L}_{\psi_{(D)}\mathcal{D}^{(*)}\mathcal{D}^{(*)}}=
		&ig_{\psi_{(D)}\mathcal{DD}}\psi^{\mu}(\partial_{\mu}\mathcal{D}^{\dagger}\mathcal{D}-\mathcal{D}^{\dagger}\partial_{\mu}\mathcal{D})  \\\nonumber
		&+g_{\psi_{(D)}\mathcal{DD}^{*}}\varepsilon_{\mu\nu\alpha\beta}\partial^{\mu}\psi^{\nu}(\mathcal{D}^{*\alpha\dagger}\overleftrightarrow{\partial}^\beta\mathcal{D}-\mathcal{D}^{\dagger}\overleftrightarrow{\partial}^\beta\mathcal{D}^{*\alpha})  \\\nonumber
		&+ig_{\psi_{(D)}{\mathcal D}^{*}{\mathcal D}^{*}}\psi^{\mu}(\partial_{\nu}{\mathcal D}_{\mu}^{*\dagger}{\mathcal D}^{*\nu}-{\mathcal D}^{*\nu\dagger}\partial_{\nu}{\mathcal D}_{\mu}^{*}\\\nonumber
		&+4{\mathcal D}^{*\nu\dagger}\partial_{\mu}{\mathcal D}_{\nu}^{*}), \\
		\mathcal{L}_{\mathcal{D}^{(*)}\mathcal{D}^{*}\eta}=
		&ig_{\mathcal{DD}^*\eta}(\mathcal{D}_{\mu}^{*\dagger}\mathcal{D}-\mathcal{D}^{\dagger}\mathcal{D}_{\mu}^*)\partial^{\mu}\eta  \\\nonumber
		&-g_{\mathcal{D}^{*}\mathcal{D}^{*}\eta}\varepsilon_{\mu\nu\alpha\beta}\partial^{\mu}\mathcal{D}^{*\dagger\nu}\partial^{\alpha}\mathcal{D}^{*\beta}\eta
	\end{align}
with $\mathcal{D}^{(*)\dagger}=(D^{(*)0},D^{(*)+})$ and $\mathcal{D}^{(*)}=(\bar{D}^{(*)0},\bar{D}^{(*)-})^T$.

	With the above preparations, we can write down the amplitudes
	of $\psi(4040)$ and $\psi(4160)$ decays into $\eta J/\psi$  within the $3S$-$2D$ mixing scheme defined in Eq. (\ref{mix}), as shown in Fig. \ref{fig: hadron loop},

\begin{eqnarray}
\mathcal{M}_{\psi{(4040)}}^{\mathrm{Total}}&=&4\sum_{\mathrm{i=a}}^{\mathrm{f}}\mathcal{M}^{(i)}_{\psi{(4040)}},\\ 
\mathcal{M}_{\psi{(4160)}}^{\mathrm{Total}}&=&4\sum_{\mathrm{i=a}}^{\mathrm{f}} {\mathcal{M}}^{(i)}_{\psi{(4160)}},
\end{eqnarray}
	where the fourfold factor comes from the charge conjugation transformation $(D^{(*)}\leftrightarrow\bar{D}^{(*)})$ and the isospin transformations $(D^{(*)0}\leftrightarrow D^{(*)+}\mathrm{~and~}\bar{D}^{(*)0}\leftrightarrow D^{(*)-})$ of the bridged $D^{(*)}$ mesons. The concrete expressions of $\mathcal{M}^{(i)}_{\psi(4040)}$ and $\mathcal{M}^{(i)}_{\psi(4160)}$ are given in Appendix \ref{amp}, the amplitudes of the others are similar to the group of $\psi(4040)$ and $\psi(4160)$ with different coupling constants and mixing angles.

	\begin{table}[htbp]
		\centering
		\caption{The coupling constants of the $g_{\psi D^{(*)}D^{(*)}}$ and $\mathcal{S}$ values  recommended in Table III of Ref. \cite{Wang:2022jxj}, where $g_{\psi(4040)D^{(*)}D^{(*)}}$ and $g_{\psi(4160)D^{(*)}D^{(*)}}$ are determined by the theoretical partial widths of $\psi(4040)/\psi(4160)\to D^{(*)}\bar{D}^{(*)}$ listed in Table \ref{width}.}
		\label{coupling Constants1}
		\renewcommand\arraystretch{1.3}
		\begin{tabular*}{86mm}{l@{\extracolsep{\fill}}ccccc}
			\toprule[1pt]
			\toprule[0.5pt]
			Charmonia  &$\theta$ &$g_{\psi DD}$ &$g_{\psi DD^*}\,(\text{GeV}^{-1})$ &$g_{\psi D^*D^*}$ &$\mathcal{S}$\\
			\midrule[0.5pt]
			$\psi(4040)$ &$20^\circ$ &0.378 &0.462 &8.607 &1.140\\
			$\psi(4160)$ &$20^\circ$ &$-2.730$ &$-0.070$ &$-1.881$ &$-0.762$\\
			$\psi(4220)$ &$34^\circ$ &0.760 &0.054 &1.220 &1.471\\
			$\psi(4380)$ &$34^\circ$ &0.570 &$-0.150$ &$-0.410$ &$-1.077$\\
			$\psi(4415)$ &$30^\circ$ &0.680 &0.003 &0.430 &1.705\\
			$\psi(4500)$ &$30^\circ$ &0.440 &$-0.076$ &$-0.015$ &$-33.989$\\
			\bottomrule[0.5pt]
			\bottomrule[1pt]
		\end{tabular*}
	\end{table}

	\begin{table}[htbp]
		\centering
		\caption{The partial width of $\psi(4040)/\psi(4160)\to D^{(*)}\bar{D}^{(*)}$ calculated by the quark pair creation (QPC) model, the parameters are in agreement with Ref. \cite{Wang:2019mhs}.}
		\label{width}
		\renewcommand\arraystretch{1.3}
		\begin{tabular*}{86mm}{l@{\extracolsep{\fill}}cccc}
			\toprule[1pt]
			\toprule[0.5pt]
			Charmonia &$D\bar{D}\,(\text{MeV})$  &$D\bar{D}^*+c.c.\,(\text{MeV})$ &$D^*\bar{D}^*\,(\text{MeV})$\\
			\midrule[0.5pt]
			$\psi(4040)$ &0.433 &17.187 &40.893\\
			$\psi(4160)$ &35.620 &0.919 &51.223\\
			\bottomrule[0.5pt]
			\bottomrule[1pt]
		\end{tabular*}
	\end{table}

	\begin{table*}[htbp]
		\renewcommand\arraystretch{1.5}
		\caption{The fitting parameters in scheme I, and the solutions $A$, $B$ and $C$ in scheme II. The $g$ and $a$ are parameters for the background, and $\mathcal{BR}_i$ and $\phi_i~\text{(rad)}$ are decay branching ratios into $J/\psi \eta$  and phases for five $\psi$ states [successively $\psi(4040)$, $\psi(4160)$, $\psi(4220)$, $\psi(4380)$ and $\psi(4415)$], respectively. }
		\label{tab:fitting parameter}
		\centering 
		\begin{tabular}{c|c|ccc}
			\hline \hline
			Parameters & Scheme I &\multicolumn{3}{c}{Scheme II} \\ \cline{3-5} 
			
			&     Fit I          ~~&~~ Solution $A$ ~~&~~ Solution $B$ ~~&~~ Solution $C$\\
			\hline 	
			$g\,(\text{GeV}^{-3})$ ~~&~~$(2.877\pm0.093)\times10^{-3}$ ~~&~~$(5.774\pm0.294)\times10^{-4}$    ~~&~~~ $(5.502\pm0.278)\times10^{-4}$  ~~&~~$(5.208\pm0.295)\times10^{-4}$\\			
			$a\,(\text{GeV}^{-2})$  ~~&~~$6.421\pm0.091$ ~~&~~$3.676\pm0.075$    ~~&~~~$3.686\pm0.082$ ~~&~~$3.628\pm0.089$\\	
			\hline
			$\mathcal{BR}_1$ ~~&~~$5.086\times10^{-4}$ ~~&~~$(2.637\pm0.213)\times10^{-3}$	~~&~~$(6.798\pm0.333)\times10^{-3}$	~~&~~$(3.336\pm0.229)\times10^{-3}$\\			
			$\mathcal{BR}_2$  ~~&~~$1.516\times10^{-3}$ ~~&~~$(2.621\pm0.154)\times10^{-3}$ 	~~&~~$(7.017\pm0.275)\times10^{-3}$ ~~&~~$(9.837\pm0.426)\times10^{-3}$\\		
			$\mathcal{BR}_3$  ~~&~~$1.970\times10^{-3}$ ~~&~~$(4.336\pm0.181)\times10^{-3}$ 	~~&~~$(5.409\pm0.214)\times10^{-3}$ ~~&~~$(2.344\pm0.053)\times10^{-2}$\\	
			$\mathcal{BR}_4$  ~~&~~$4.063\times10^{-4}$ ~~&~~$(1.357\pm0.121)\times10^{-3}$ 	~~&~~$(1.620\pm0.158)\times10^{-3}$ ~~&~~$(2.074\pm0.188)\times10^{-3}$\\	
			$\mathcal{BR}_5$  ~~&~~$1.746\times10^{-3}$ ~~&~~$(2.979\pm0.560)\times10^{-4}$	~~&~~$(3.038\pm0.638)\times10^{-4}$ ~~&~~$(3.143\pm0.620)\times10^{-4}$\\
			\hline 	
			$\phi_1~\text{(rad)}$  ~~&~~$2.346\pm0.103$ ~~&~~$3.087\pm0.050$	~~&~~$3.899\pm0.030$  ~~&~~$3.539\pm0.045$\\			
			$\phi_2~\text{(rad)}$  ~~&~~$0.666\pm0.067$ ~~&~~$2.710\pm0.050$	~~&~~$1.737\pm0.026$  ~~&~~$3.879\pm0.018$\\	
			$\phi_3~\text{(rad)}$  ~~&~~$2.413\pm0.060$ ~~&~~$4.404\pm0.057$	~~&~~$3.750\pm0.056$  ~~&~~$1.475\pm0.016$\\		
			$\phi_4~\text{(rad)}$  ~~&~~$5.107\pm0.080$ ~~&~~$0.298\pm0.069$	~~&~~$0.190\pm0.060$  ~~&~~$6.212\pm0.054$\\		
			$\phi_5~\text{(rad)}$  ~~&~~$4.263\pm0.081$ ~~&~~$0.243\pm0.230$	~~&~~$0.161\pm0.223$  ~~&~~$6.217\pm0.217$\\			
			\hline 		
			$\chi^2/\text{d.o.f.}$  ~~&~~1.492 ~~&~~ 0.938	~~&~~ 0.935 ~~&~~ 0.931 \\
			\hline \hline 	
		\end{tabular}
	\end{table*}
	

	Finally, the branching ratio of the charmonium decay into $\eta J/\psi$ can be obtained by
	\begin{align}
		\mathcal{BR}[\psi\to J/\psi \eta]=\frac13\frac{|\vec{p}_1|}{8\pi m_{\psi}^2}\sum_{spin}\Big|{\mathcal{M}_\psi}^{\mathrm{Total}}\Big|^2\Big/\Gamma_{\psi},
	\end{align}
	where $m_{\psi}$ and $\Gamma_{\psi}$ are the mass and total width of the initial charmonium listed in Table \ref{tab: parameter}, and $\vec{p}_1$ is the three-momentum of the $\eta$ meson in the rest frame of the initial state. The coefficient 1/3 and the sum $\displaystyle\sum_{spin}$ come from the averaging over the polarizations of the initial state and summing up the polarizations of the final state.

	Below is a brief description of the method for determining the relevant coupling constants, as they appear in the concrete amplitudes outlined in Appendix \ref{amp}.

	The coupling constants of $g_{D^{(*)}D^*\eta}$ are given in Ref. \cite{Li:2021jjt},
	\begin{align}
		\frac{g_{D^*D\eta}}{\sqrt{m_Dm_{D^*}}}=g_{D^*D^*\eta}=\frac{2g_H}{f_\pi}\alpha_\theta  ,
	\end{align}
	within $\alpha_\theta=\frac{1}{\sqrt 6}\cos{(-19.1^\circ)}$ and ${f_\pi}=131~\text{MeV}$, and $g_H=0.569$ is determined by the measured decay width of $D^{*+}\to D^0\pi^+$ \cite{Wang:2016qmz}. $g_{J/\psi DD}=7.44$, $g_{J/\psi DD^*}=3.84~\text{GeV}^{-1}$ and $g_{J/\psi D^*D^*}=8.00$ are obtained by the vector meson dominance model \cite{Achasov:1994vh,Deandrea:2003pv}.
	
	
	When taking the $3S$-$2D$ mixing charmonia $\psi(4040)$ and $\psi(4160)$ as examples, the coupling constants of $g_{\psi(4040)D^{(*)}D^{(*)}}$ and $g_{\psi(4160)D^{(*)}D^{(*)}}$ are defined by
	\begin{equation}
	\begin{aligned}
	g_{\psi(4040) D^{(*)}D^{(*)}}&=g_{\psi(3S)D^{(*)}D^{(*)}}\cos\theta+g_{\psi(2D)D^{(*)}D^{(*)}}\sin\theta \\
	g_{\psi(4160) D^{(*)}D^{(*)}}&=-g_{\psi(3S)D^{(*)}D^{(*)}}\sin\theta+g_{\psi(2D)D^{(*)}D^{(*)}}\cos\theta,
	\end{aligned}
	\end{equation}
	The values are determined by the theoretical partial widths of $\psi(4040)/\psi(4160)\to D^{(*)}D^{(*)}$ calculated by the QPC model \cite{Jacob:1959at,LeYaouanc:1977gm,Micu:1968mk}, in which the mixing angle $\theta=20^\circ$ is considered \cite{Wang:2022jxj}. The other parameters required by the model are in agreement with Ref. \cite{Wang:2019mhs}, and the calculated partial widths associated with $\psi(4040)$ and $\psi(4160)$ are listed in Table \ref{width}. Other coupling constants of $g_{\psi(4220)D^{(*)}D^{(*)}}$, $g_{\psi(4380)D^{(*)}D^{(*)}}$, $g_{\psi(4415)D^{(*)}D^{(*)}}$ and $g_{\psi(4500)D^{(*)}D^{(*)}}$ have been calculated in Table III of Ref. \cite{Wang:2022jxj}, which were determined by the corresponding experimental or theoretical partial widths of $\psi\to D^{(*)}D^{(*)}$. Both of them are listed in Table \ref{coupling Constants1}.
	In addition, for convenience, the $\mathcal{S}$ factors in the decay amplitudes of Appendix \ref{amp} are defined as
	\begin{equation}
	\begin{aligned}
	\mathcal{S}_{\psi(4040)}&=\frac{g_{\psi(3S)D^{*}D^{*}}\cos\theta+4g_{\psi(2D)D^{*}D^{*}}\sin\theta}{g_{\psi(3S)D^{*}D^{*}}\cos\theta+g_{\psi(2D)D^{*}D^{*}}\sin\theta} \\
	\mathcal{S}_{\psi(4160)}&=\frac{-g_{\psi(3S)D^{*}D^{*}}\sin\theta+4g_{\psi(2D)D^{*}D^{*}}\cos\theta}{-g_{\psi(3S)D^{*}D^{*}}\sin\theta+g_{\psi(2D)D^{*}D^{*}}\cos\theta}.
	\end{aligned}
	\end{equation}

	\begin{figure}[tph]
		\centering
		\includegraphics[width=0.45\textwidth]{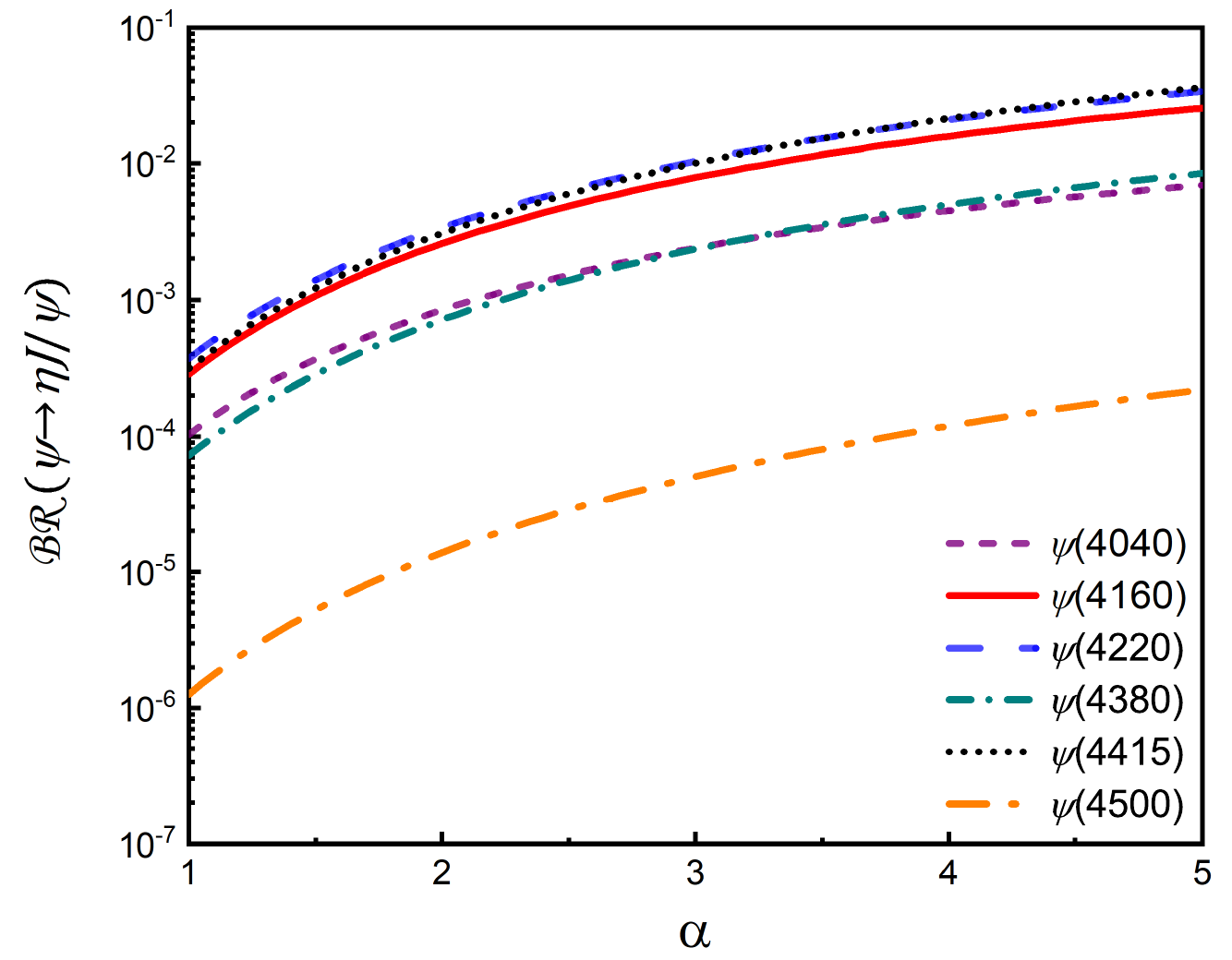}
		\caption{The $\alpha$ parameter dependence of the predicted branching ratios, including $\mathcal{BR}(\psi(4040) \to \eta J/\psi)$, $\mathcal{BR}(\psi(4160) \to \eta J/\psi)$, $\mathcal{BR}(\psi(4220) \to \eta J/\psi)$, $\mathcal{BR}(\psi(4380) \to \eta J/\psi)$,  $\mathcal{BR}(\psi(4415) \to \eta J/\psi)$, and $\mathcal{BR}(\psi(4500) \to \eta J/\psi)$.}
		\label{fig: branching ratio}
	\end{figure}
	
	
With the determined coupling constants discussed above, we calculate the branching ratios of the six theoretically constructed charmonia, i.e., $\psi(4040)$, $\psi(4160)$, $\psi(4220)$, $\psi(4380)$, $\psi(4415)$ and $\psi(4500)$ decays into $\eta J/\psi$, which depend on the cutoff parameter $\alpha$ introduced by the form factor of Eq. (\ref{FFs}) in the range of $[1,5]$, as shown in Fig. \ref{fig: branching ratio}. It can be seen that the branching ratios associated with $\psi(4040)$, $\psi(4160)$, $\psi(4220)$ and $\psi(4380)$, $\psi(4415)$ can reach the order of $10^{-4}\rm{-}10^{-2}$ and $10^{-4}\rm{-}10^{-3}$, respectively, while $\mathcal{BR}(\psi(4500) \to \eta J/\psi)$ is only of order of $10^{-6}\rm{-}10^{-5}$.

	\subsection{A combined fit to the cross section data of $e^+e^- \to \eta J/\psi$}\label{SubB}
	
	With the input of the branching ratios $\mathcal{BR}(\psi\to \eta J/\psi)$  predicted by the hadronic loop mechanism as shown in Fig. \ref{fig: branching ratio}, we can now obtain the contribution of each vector charmonium to the cross section of $e^+e^- \to  \eta J/\psi$ from Eq. (\ref{a}), and apply a combined fit to the experimental data.
	
	Since the branching ratio of $\psi(4500)\to \eta J/\psi$ is at least 2 orders of magnitude smaller than the others, and there is indeed no signal of structures around 4.5 GeV in the measured data of $e^+e^- \to \eta J/\psi$, we believe that the contribution of $\psi(4500)$ to the cross section of $e^+e^- \to \eta J/\psi$ can be safely dropped. In the following, we will use the left five theoretically constructed charmonia, i.e., $\psi(4040)$, $\psi(4160)$, $\psi(4220)$, $\psi(4380)$, and $\psi(4415)$, as well as their theoretical branching ratios of decays into $\eta J/\psi$, dependent on the $\alpha$ parameter, to perform a combined fit to the cross section data of $e^+e^- \to \eta J/\psi$.

	\begin{figure}[htbp]\centering
		\includegraphics[width=0.47\textwidth]{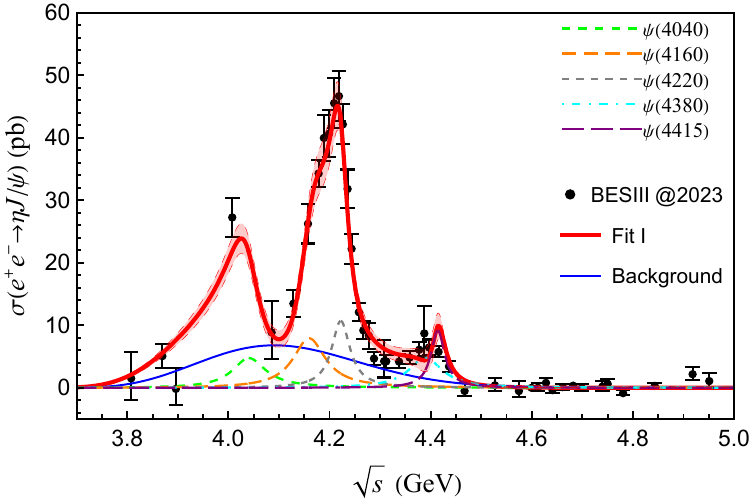}
		\caption{ Our fit to the higher vector charmonium contribution in the cross section distribution of $e^+e^- \to \eta J/\psi$ within scheme I. Here, the data points are from BESIII measurement~\cite{BESIII:2023tll}, the five dashed lines represent the contributions of the higher charmonium states, the blue line represents the background, and the red line with a band represents the total contribution and uncertainties.}
		\label{fig:fit1}
	\end{figure}
	
	\begin{figure*}[htbp]
	\centering
		\includegraphics[width=0.47\textwidth]{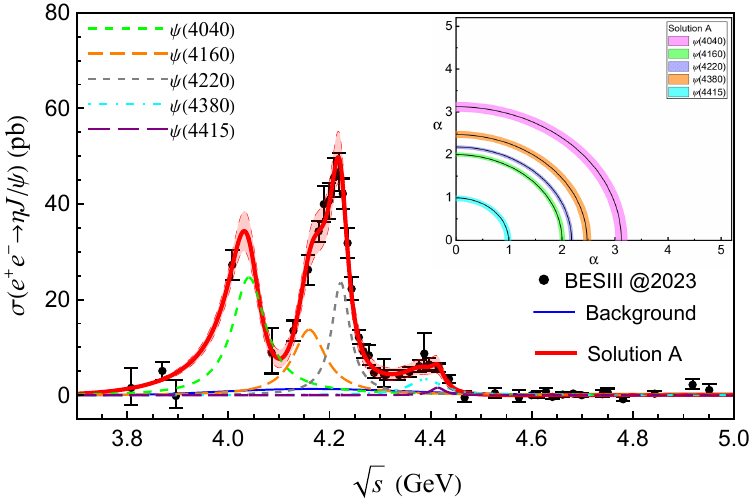}\\
		\includegraphics[width=0.47\textwidth]{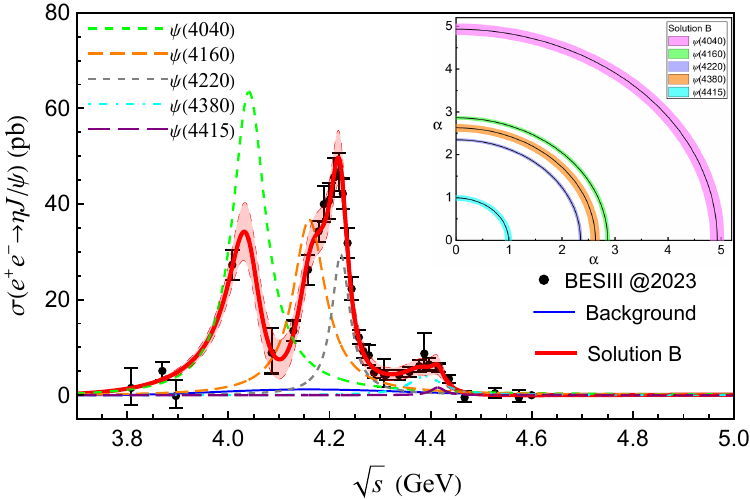}\\	
		\includegraphics[width=0.47\textwidth]{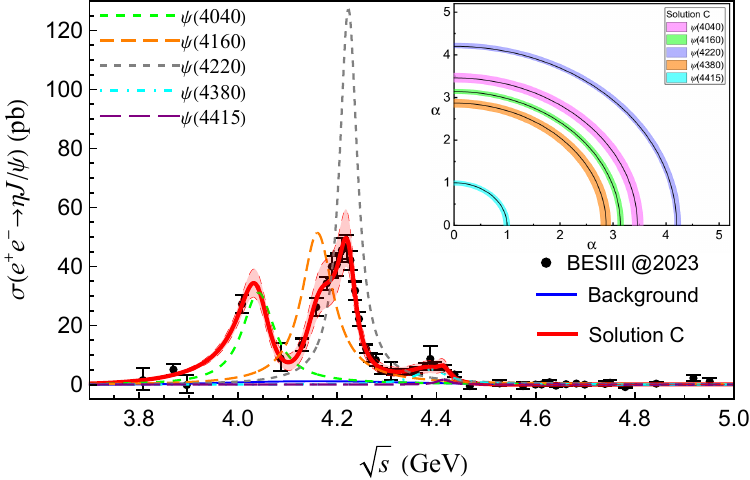}\\
		\caption{Our fit to the cross section of the  $e^+e^- \to \eta J/\psi$ process between $E_{cm}=3.808$ to 4.600 GeV by scheme II. Here, the five dashed lines represent the contributions of the higher charmonium states, the blue line represents the background, and the red line with a band represents the total contribution and uncertainties. The insets show the branching ratios for each state with the central values and uncertainties in the fit as shown in Table~\ref{tab:fitting parameter}  and are represented by the corresponding $\alpha$ values in the hadronic loop mechanism [in Eq.~(\ref{FFs})] as shown in Fig.~\ref{fig: branching ratio}, where the central values and the errors are represented by the solid black lines and the colored bands, respectively.
		}
		\label{fig:fit2}
	\end{figure*}

	The total amplitude of $e^+e^- \to \eta J/\psi$ can be written as	
	\begin{align}
		\mathcal{M}_{\mathrm{Total}}(s)=\mathcal{M}_{0}(s)+\sum_ie^{i\phi_i}\mathcal{M}_{\psi_i}(s),
	\end{align}
	with an exponentially parametrized background formulated as
	\begin{align}
		\mathcal{M}_{0}(s)=g(\sqrt{s}-m_\eta-m_{J/\psi})^2e^{-a(\sqrt{s}-m_\eta-m_{J/\psi})^2},
	\end{align}
	where $\phi_i$ is the phase angle between the resonance amplitude $\mathcal{M}_{\psi_i}(s)$ and the background $\mathcal{M}_{0}(s)$, $g$ and $a$ are two free parameters. 
	
	The total cross section can be represented by
	\begin{align}
	\label{fitfunction}
		\sigma(s)=|\mathcal{M}_{\mathrm{Total}}(s)|^{2}.
	\end{align}

	Here, we employ two fitting schemes. In the first scheme, we assume a same $\alpha$ parameter for all charmonium states. The best fitting result in scheme I, represented by a red curve, is shown in Fig. \ref{fig:fit1}, with $\chi^2/\text{d.o.f.}=1.49$. The relevant fitting parameters in scheme I are listed in Table \ref{tab:fitting parameter}.
	
	A direct observation in Fig. \ref{fig:fit1} shows that the line shape around 4.0 GeV is not well reproduced in scheme I. This is mainly due to the scarcity of cross section data points around 4.0 GeV, which puzzles us in determining the contribution of the resonance state $\psi(4040)$. Our analysis suggests two resonance contributions from $\psi(4160)$ and $\psi(4220)$ around 4.2 GeV. 
	With the same $\alpha$ parameter, the branching ratio of $\psi(4415)$ is significantly larger than that of $\psi(4380)$ (see Fig. \ref{fig: branching ratio}), thus the peak around 4.4 GeV is dominated by the contribution of $\psi(4415)$, which affects the quality of the fit and is one of the largest contributions to the $\chi^2/\text{d.o.f.}$. Therefore, it should be a focus of future BESIII and Belle II experiments to have a more clear understanding around 4.4 GeV in this reaction process. In particular, we found that the interference effect plays a crucial role in broadening the distribution of the resonance signal associated with $Y(4230)$, explaining its large width of $80.4\pm4.4\pm1.4$ MeV in the experimental fit and the line shape puzzle around 4.2 GeV in the $e^+e^- \to \eta J/\psi$ cross section.
	
	In scheme II, we consider a range of $\alpha$ parameters $[1,5]$ for the five charmonium states without extra limit. The three solutions $A$, $B$, and $C$ in scheme II, represented by the red curve, are shown in Fig. \ref{fig:fit2} with $\chi^2/\text{d.o.f.}$ around 0.93, which is obviously improved compared to that of scheme I. The relevant fitting parameters in scheme II are listed in Table \ref{tab:fitting parameter}.
	
	Similar to scheme I, our analysis in scheme II shows the significant role of the interference effects in broadening the distribution of the resonance signal associated with $\psi(4160)$ and $\psi(4220)$, which naturally explains the asymmetric line shape associated with the resonance structure denoted as $Y(4230)$. Of the three solutions, $\psi(4040)$ gives different contributions to the cross section, and we hope that the experiment will complement the data points around in this part and clarify the contribution of $\psi(4040)$. Around 4.4 GeV, each solution in scheme II shows an unremarkable resonance signal of $\psi(4415)$ in $e^+e^- \to \eta J/\psi$. Further investigation in future BESIII and Belle II experiments is indeed necessary to solve the broad width puzzle associated with the resonance structure denoted as $Y(4360)$.

	\section{SUMMARY}\label{SecIII}

	Recently, the BESIII collaboration measured the cross section of $e^+e^-\to \eta J/\psi$  from 3.808 to 4.951 GeV, and reported three structures, the charmonium $\psi(4040)$ and two other charmoniumlike states, named $Y(4230)$ and $Y(4360)$ \cite{BESIII:2023tll}. It is curious that the measured width of $Y(4230)$ is so large and has a large difference from the other open-charm and hidden-charm measurements, and the more subtle details of the line shape around 4.2 GeV with asymmetry suggest that the structure of the observed $Y(4230)$ may not be formed by a single resonance. 
	
	In order to understand the puzzles arising in the measured data, the energy level structures of the charmonia from theoretical inputs are crucial. In this work, we focus on the six charmonia constructed by an unquenched potential model in the range of $4.0\rm{-}4.5$ GeV, i.e., $\psi(4040)$, $\psi(4160)$, $\psi(4220)$, $\psi(4380)$, $\psi(4415)$, and $\psi(4500)$. Using the hadronic loop mechanism, we are able to quantitatively calculate the branching ratios of the higher charmonium state decays into $\eta J/\psi$. The results indicate that the contribution of $\psi(4500)$ to the cross section of $e^+e^-\to \eta J/\psi$ is too small and is ignored in our later fitting analysis, which is in agreement with the measured data, while others have sizable contributions. We then perform a combined fit with the left five charmonia to the newly measured cross section data of $e^+e^-\to \eta J/\psi$, the corresponding branching ratios of these charmonia decaying into $\eta J/\psi$ are constrained within a reasonable range, suggested by the hadronic loop mechanism. We have made two attempts with different cutoff parameter constraint schemes. Both fitting results show that the newly measured cross section of $e^+e^-\to \eta J/\psi$ can be reproduced by the five charmonia from the theoretical inputs, the puzzling large width and asymmetric line shape of the 4.2 GeV structure can be naturally explained by the contributions of the neighboring charmonia $\psi(4160)$ and $\psi(4220)$. Moreover, the introduction of $\psi(4380)$ predicted by the theory in the combined fit is compatible with the experimental data, which together with $\psi(4415)$ can reproduce the third broad enhancement structure around 4.4 GeV reported by BESIII.  More importantly, the results support the characterized energy level construction of higher vector charmonia in the range of $4.0\rm{-}4.5$ GeV.
	
	With the increasing accumulation of experimental data on the varieties of final states in this particular charmonium energy range, as well as the continuing attention of experimentalists and theorists, our understanding of the inner nature of higher charmonium states is becoming more mature and profound.

	\section*{ACKNOWLEDGMENTS}
	
	This work is supported by the National Natural Science Foundation of China under Grant Nos. 12335001 and 12247101, the China National Funds for Distinguished Young Scientists under Grant No. 11825503, National Key Research and Development Program of China under Contract No. 2020YFA0406400, the 111 Project under Grant No. B20063, the fundamental Research Funds for the Central Universities, and the project for top-notch innovative talents of Gansu province. J.Z.W. is also supported by the National Postdoctoral Program for Innovative Talent.

	\appendix
	\begin{widetext}
		\section{The decay amplitudes of hadronic loop mechanism} \label{amp}
		\begin{align}
			\mathcal{M}_{\psi(4040)}^{(a)}=
			&\int \frac{d^4q}{(2\pi)^4} g_{\psi(4040)DD}\times \epsilon_{\psi}^{\mu}(q_{2\mu}-q_{1\mu})(-g_{D^{*}D\eta})p_{1}^{\alpha}g_{J/\psi DD^{*}}\varepsilon_{\lambda\gamma\beta\rho}p_{2}^{\lambda}\epsilon_{J/\psi}^{\gamma*}(q_{2}^{\rho}-q^{\rho})\\\nonumber
			&\times \left(-g_{\alpha}^{\beta}+q_{\alpha}q^{\beta}/m_{D^{*}}^{2}\right)\frac{\mathcal{F}^{2}(q^{2},m_{B^{*}}^{2})}{{q^{2}-m_{D^{*}}^{2}}}, \\\nonumber
		\end{align}
		\begin{align}
			\mathcal{M}_{\psi(4040)}^{(b)}=&\int \frac{d^4q}{(2\pi)^4} g_{\psi(4040)DD^*}\varepsilon_{\nu\mu\alpha\beta}\epsilon_{\psi}^{\mu}(q_2^\beta-q_1^\beta)(-g_{D^{*}D\eta})p_1^\gamma (g_{J/\psi D^*D^*})\epsilon_{J/\psi}^{\rho*}[g_{\sigma\rho}q_{2\eta}-g_{\eta\rho}q_{1\sigma}+g_{\theta\eta}(q_\rho-q_{2\rho})]\\\nonumber
			&\times\left(-g^{\alpha\sigma}+q_2^{\alpha}q_2^{\sigma}/m_{D^{*}}^{2}\right)\left(-g_\gamma^\eta+q_\gamma q^\eta/m_{D^{*}}^{2}\right)	
			\frac{\mathcal{F}^{2}(q^{2},m_{B^{*}}^{2})}{{q^{2}-m_{D^*}^{2}}}, \\\nonumber
		\end{align}
		\begin{align}
			\mathcal{M}_{\psi(4040)}^{(c)}=&\int \frac{d^4q}{(2\pi)^4} g_{\psi(4040)DD^*}\varepsilon_{\nu\mu\alpha\beta}\epsilon_{\psi}^{\mu}p^\nu(q_2^\beta-q_1^\beta)(g_{D^{*}D\eta})p_1^\gamma(g_{J/\psi DD})\epsilon_{J/\psi}^{\rho*}(q_\rho-q_{2\rho})\\\nonumber
			&\times \left(-g^\alpha_\gamma+q_1^{\alpha}q_1^{\gamma}/m_{D^{*}}^{2}\right)\frac{\mathcal{F}^{2}(q^{2},m_{B^{*}}^{2})}{{q^{2}-m_{D}^{2}}}, \\\nonumber
		\end{align}
		\begin{align}
			\mathcal{M}_{\psi(4040)}^{(d)}=&\int \frac{d^4q}{(2\pi)^4} g_{\psi(4040)DD^*}\varepsilon_{\nu\mu\alpha\beta}\epsilon_{\psi}^{\mu}(q_2^\beta-q_1^\beta)(-g_{D^{*}D^{*}\eta})\varepsilon_{\rho\eta\lambda\theta}q^\rho q_1^\lambda(g_{J/\psi DD^{*}})\varepsilon_{\delta\theta\tau\gamma}\epsilon_{J/\psi}^{\theta*}p_{2}^{\delta}(q^\gamma-q_2^\gamma)\\\nonumber
			&\times\left(-g^{\alpha\eta}+q_1^{\alpha}q_1^{\eta}/m_{D^{*}}^{2}\right)\left(-g^{\tau\sigma}+q^{\tau}q^{\sigma}/m_{D^{*}}^{2}\right)
			\frac{\mathcal{F}^{2}(q^{2},m_{B^{*}}^{2})}{{q^{2}-m_{D^{*}}^{2}}}, \\\nonumber
		\end{align}
		\begin{align}
			\mathcal{M}_{\psi(4040)}^{(e)}=&\int \frac{d^4q}{(2\pi)^4} g_{\psi(4040)D^*D^*} \epsilon_{\psi}^{\mu}[g_{\beta\mu}q_{2\alpha}-g_{\alpha\mu}q_{1\beta}+\mathcal{S}\cdot g_{\alpha\beta}(q_{1\mu}-q_{2\mu})] (g_{D^{*}D\eta})p_1^\gamma (g_{J/\psi DD^{*}})\varepsilon_{\lambda\eta\sigma\rho}\epsilon_{J/\psi}^{\eta*}p_{2}^{\lambda}(q^\rho-q_2^\rho)\\\nonumber
			&\times \left(-g_\gamma^{\alpha}+q_1^{\alpha}q_{1\gamma}/m_{D^{*}}^{2}\right)\left(-g^{\beta\sigma}+q^\beta q^\sigma/m_{D^{*}}^{2}\right) \frac{\mathcal{F}^{2}(q^{2},m_{B^{*}}^{2})}{{q^{2}-m_{D^*}^{2}}},	\\\nonumber
		\end{align}
		\begin{align}
			\mathcal{M}_{\psi(4040)}^{(f)}=&\int \frac{d^4q}{(2\pi)^4}g_{\psi(4040)D^*D^*}\epsilon_{\psi}^{\mu}[g_{\beta\mu}q_{2\alpha}-g_{\alpha\mu}q_{1\beta}+\mathcal{S}\cdot g_{\alpha\beta}(q_{1\mu}-q_{2\mu})](-g_{D^{*}D^*\eta})\varepsilon_{\rho\sigma\lambda\eta}q_1^\lambda q^\rho (g_{J/\psi D^{*}D^{*}})\\\nonumber
			&\times\epsilon_{J/\psi}^{\gamma*}[g_{\tau\gamma}q_{2\theta}-g_{\theta\gamma}q_{1\tau}+g_{\tau\theta}(q_\gamma-q_{2\gamma})]\\\nonumber	
			&\times \left(-g^{\alpha\eta}+q_1^{\alpha}q_1^{\eta}/m_{D^{*}}^{2}\right)\left(-g^{\tau\beta}+q_2^{\tau}q_2^{\beta}/m_{D^{*}}^{2}\right) \left(-g^{\sigma\theta}+q^\sigma q^\theta/m_{D^{*}}^{2}\right)\frac{\mathcal{F}^{2}(q^{2},m_{B^{*}}^{2})}{{q^{2}-m_{D^*}^{2}}}, \\\nonumber
		\end{align}
		\begin{align}
			\mathcal{M}_{\psi(4160)}^{(a)}=
			&\int \frac{d^4q}{(2\pi)^4} g_{\psi(4160)DD}\times \epsilon_{\psi}^{\mu}(q_{2\mu}-q_{1\mu})(-g_{D^{*}D\eta})p_{1}^{\alpha}g_{J/\psi DD^{*}}\varepsilon_{\lambda\gamma\beta\rho}p_{2}^{\lambda}\epsilon_{J/\psi}^{\gamma*}(q_{2}^{\rho}-q^{\rho})\\\nonumber
			&\times \left(-g_{\alpha}^{\beta}+q_{\alpha}q^{\beta}/m_{D^{*}}^{2}\right)\frac{\mathcal{F}^{2}(q^{2},m_{B^{*}}^{2})}{{q^{2}-m_{D^{*}}^{2}}}, \\\nonumber
		\end{align}
		\begin{align}
			\mathcal{M}_{\psi(4160)}^{(b)}=&\int \frac{d^4q}{(2\pi)^4} g_{\psi(4160)DD^*}\varepsilon_{\nu\mu\alpha\beta}\epsilon_{\psi}^{\mu}(q_2^\beta-q_1^\beta)(-g_{D^{*}D\eta})p_1^\gamma (g_{J/\psi D^*D^*})\epsilon_{J/\psi}^{\rho*}[g_{\sigma\rho}q_{2\eta}-g_{\eta\rho}q_{1\sigma}+g_{\theta\eta}(q_\rho-q_{2\rho})]\\\nonumber
			&\times\left(-g^{\alpha\sigma}+q_2^{\alpha}q_2^{\sigma}/m_{D^{*}}^{2}\right)\left(-g_\gamma^\eta+q_\gamma q^\eta/m_{D^{*}}^{2}\right)	
			\frac{\mathcal{F}^{2}(q^{2},m_{B^{*}}^{2})}{{q^{2}-m_{D^*}^{2}}}, \\\nonumber
		\end{align}
		\begin{align}
			\mathcal{M}_{\psi(4160)}^{(c)}=&\int \frac{d^4q}{(2\pi)^4} g_{\psi(4160)DD^*}\varepsilon_{\nu\mu\alpha\beta}\epsilon_{\psi}^{\mu}p^\nu(q_2^\beta-q_1^\beta)(g_{D^{*}D\eta})p_1^\gamma(g_{J/\psi DD})\epsilon_{J/\psi}^{\rho*}(q_\rho-q_{2\rho})\\\nonumber
			&\times \left(-g^\alpha_\gamma+q_1^{\alpha}q_1^{\gamma}/m_{D^{*}}^{2}\right)\frac{\mathcal{F}^{2}(q^{2},m_{B^{*}}^{2})}{{q^{2}-m_{D}^{2}}}, \\\nonumber
		\end{align}
		\begin{align}
			\mathcal{M}_{\psi(4160)}^{(d)}=&\int \frac{d^4q}{(2\pi)^4} g_{\psi(4160)DD^*}\varepsilon_{\nu\mu\alpha\beta}\epsilon_{\psi}^{\mu}(q_2^\beta-q_1^\beta)(-g_{D^{*}D^{*}\eta})\varepsilon_{\rho\eta\lambda\theta}q^\rho q_1^\lambda(g_{J/\psi DD^{*}})\varepsilon_{\delta\theta\tau\gamma}\epsilon_{J/\psi}^{\theta*}p_{2}^{\delta}(q^\gamma-q_2^\gamma)\\\nonumber
			&\times\left(-g^{\alpha\eta}+q_1^{\alpha}q_1^{\eta}/m_{D^{*}}^{2}\right)\left(-g^{\tau\sigma}+q^{\tau}q^{\sigma}/m_{D^{*}}^{2}\right)
			\frac{\mathcal{F}^{2}(q^{2},m_{B^{*}}^{2})}{{q^{2}-m_{D^{*}}^{2}}}, \\\nonumber
		\end{align}
		\begin{align}
			\mathcal{M}_{\psi(4160)}^{(e)}=&\int \frac{d^4q}{(2\pi)^4} g_{\psi(4160)D^*D^*} \epsilon_{\psi}^{\mu}[g_{\beta\mu}q_{2\alpha}-g_{\alpha\mu}q_{1\beta}+\mathcal{S}\cdot g_{\alpha\beta}(q_{1\mu}-q_{2\mu})] (g_{D^{*}D\eta})p_1^\gamma (g_{J/\psi DD^{*}})\varepsilon_{\lambda\eta\sigma\rho}\epsilon_{J/\psi}^{\eta*}p_{2}^{\lambda}(q^\rho-q_2^\rho)\\\nonumber
			&\times \left(-g_\gamma^{\alpha}+q_1^{\alpha}q_{1\gamma}/m_{D^{*}}^{2}\right)\left(-g^{\beta\sigma}+q^\beta q^\sigma/m_{D^{*}}^{2}\right) \frac{\mathcal{F}^{2}(q^{2},m_{B^{*}}^{2})}{{q^{2}-m_{D^*}^{2}}},	\\\nonumber	
		\end{align}
		\begin{align}
			\mathcal{M}_{\psi(4160)}^{(f)}=&\int \frac{d^4q}{(2\pi)^4}g_{\psi(4160)D^*D^*}\epsilon_{\psi}^{\mu}\left[g_{\beta\mu}q_{2\alpha}-g_{\alpha\mu}q_{1\beta}+\mathcal{S}\cdot g_{\alpha\beta}(q_{1\mu}-q_{2\mu})\right]\left(-g_{D^{*}D^*\eta}\right)\varepsilon_{\rho\sigma\lambda\eta}q_1^\lambda q^\rho (g_{J/\psi D^{*}D^{*}})\\\nonumber
			&\times\epsilon_{J/\psi}^{\gamma*}[g_{\tau\gamma}q_{2\theta}-g_{\theta\gamma}q_{1\tau}+g_{\tau\theta}(q_\gamma-q_{2\gamma})]\\\nonumber				
			&\times \left(-g^{\alpha\eta}+q_1^{\alpha}q_1^{\eta}/m_{D^{*}}^{2}\right)\left(-g^{\tau\beta}+q_2^{\tau}q_2^{\beta}/m_{D^{*}}^{2}\right) \left(-g^{\sigma\theta}+q^\sigma q^\theta/m_{D^{*}}^{2}\right)\frac{\mathcal{F}^{2}(q^{2},m_{B^{*}}^{2})}{{q^{2}-m_{D^*}^{2}}}, 
		\end{align}
		
		\section{The correlation matrices between the parameters listed in Table \ref{tab:fitting parameter}} \label{cor}
		
	
	The correlation matrix presents the correlation between each pair of fit parameters being estimated, is defined as ${\rm{cross\rm{-}correlation}}_{ij}=\frac{(J^TJ)_{ij}^{-1}}{\sqrt{(J^TJ)_{ii}^{-1}(J^TJ)_{jj}^{-1}}}$~\cite{Michael:2020}. The $J$ matrix is defined as $J_{ij}=\frac{1}{\sigma_i}\frac{\partial\sigma(s_i,\alpha)}{\partial\alpha_j}$, where $\sigma(s_i,\alpha)$ is the fitting function defined in Eq. (\ref{fitfunction}), $s_i$ and $\sigma_i$ are the square of the center-of-mass energy and standard error of the $i$th point from the measured cross section data \cite{BESIII:2023tll}, and $\alpha_j$ is $j$th parameter as shown in Table \ref{tab:fitting parameter}.
		
		For Scheme I, the order of the fit parameters is $\{g,\,a,\,\phi_1,\,\phi_2,\,\phi_3,\,\phi_4,\,\phi_5\}$, and the correlation matrix read below,
		
	\begin{center}
		$\begin{pmatrix}\label{Scheme I}
1.000  &0.728  &-0.304  &-0.505  &-0.460  &0.112  &-0.001\\
0.728  &1.000  &-0.796  &-0.417  &-0.386  &0.527  &-0.348\\
-0.304  &-0.796  &1.000  &0.502  &-0.001  &-0.449  &0.394\\
-0.505  &-0.417  &0.502  &1.000  &0.092  &0.150  &0.297\\
-0.460  &-0.386  &-0.001  &0.092  &1.000  &-0.331  &0.128\\
0.112  &0.527  &-0.449  &0.150  &-0.331  &1.000  &-0.515\\
-0.001  &-0.348  &0.394  &0.297  &0.128  &-0.515  &1.000
\end{pmatrix}.$	    
	\end{center}

For Scheme II, the order of the fit parameters is $\{g,\,a,\,\mathcal{BR}_1,\,\mathcal{BR}_2,\,\mathcal{BR}_3,\,\mathcal{BR}_4,\,\mathcal{BR}_5,\,\phi_1,\,\phi_2,\,\phi_3,\,\phi_4,\,\phi_5\}$.
The correlation matrix of Solution $A$ is
	\begin{center}\label{Scheme IIA}
$\begin{pmatrix}
1.000 &0.943 &-0.988 &-0.936 &-0.695 &0.653 &0.644 &-0.938 &-0.968 &-0.982 &-0.612 &-0.394 \\
0.943 &1.000 &-0.912 &-0.795 &-0.438 &0.864 &0.860 &-0.991 &-0.979 &-0.951 &-0.330 &-0.079 \\
-0.988 &-0.912 &1.000 &0.969 &0.757 &-0.607 &-0.598 &0.929 &0.967 &0.991 &0.675 &0.462 \\
-0.936 & -0.795 & 0.969 & 1.000 & 0.891 & -0.405 & -0.395 & 0.820 & 0.884 & 0.937 & 0.830 & 0.658 \\
-0.695 & -0.438 & 0.757 & 0.891 & 1.000 & 0.049 & 0.059 & 0.475 & 0.580 & 0.681 & 0.991 & 0.925 \\
0.653 &0.864 &-0.607 &-0.405 &0.049 &1.000 &0.996 &-0.855 &-0.785 &-0.697 &0.172 &0.420 \\
0.644 &0.860 &-0.598 &-0.395 &0.059 &0.996 &1.000 & -0.847 &-0.776 & -0.688 & 0.182 & 0.429 \\
-0.938 & -0.991 & 0.929 & 0.820 & 0.475 & -0.855 & -0.847 & 1.000 & 0.992 & 0.968 & 0.363 & 0.110 \\
-0.968 & -0.979 & 0.967 & 0.884 & 0.580 & -0.785 & -0.776 & 0.992 & 1.000 & 0.991 & 0.476 & 0.232 \\
-0.982 & -0.951 & 0.991 & 0.937 & 0.681 & -0.697 & -0.688 & 0.968 & 0.991 & 1.000 & 0.586 & 0.357 \\
-0.612 & -0.330 & 0.675 & 0.830 & 0.991 & 0.172 & 0.182 & 0.363 & 0.476 & 0.586 & 1.000 & 0.966 \\
-0.394 & -0.079 & 0.462 & 0.658 & 0.925 & 0.420 & 0.429 & 0.110 & 0.232 & 0.357 & 0.966 & 1.000 
\end{pmatrix},$
	\end{center}
	
the correlation matrix of Solution $B$ is
	\begin{center}\label{Scheme IIB}
$\begin{pmatrix}
1.000& 0.948& -0.987& -0.934& -0.717& 0.604& 0.548& -0.942& -0.973& -0.986& -0.722& -0.565\\
0.948& 1.000& -0.920& -0.781& -0.487& 0.821& 0.779& -0.994& -0.985& -0.964& -0.480& -0.286\\
-0.987& -0.920& 1.000& 0.957& 0.780& -0.553& -0.492& 0.928& 0.966& 0.986& 0.775& 0.623\\
-0.934& -0.781& 0.957& 1.000& 0.911& -0.300& -0.234& 0.789& 0.860& 0.909& 0.919& 0.815\\
-0.717& -0.487& 0.780& 0.911& 1.000& 0.065& 0.120& 0.512& 0.614& 0.694& 0.990& 0.957\\
0.604& 0.821& -0.553& -0.300& 0.065& 1.000& 0.964& -0.819& -0.739& -0.665& 0.090& 0.300\\
0.548& 0.779& -0.492& -0.234& 0.120& 0.964& 1.000& -0.766& -0.684& -0.605& 0.146& 0.347\\
-0.942& -0.994& 0.928& 0.789& 0.512& -0.819& -0.766& 1.000& 0.992& 0.973& 0.496& 0.298\\
-0.973& -0.985& 0.966& 0.860& 0.614& -0.739& -0.684& 0.992& 1.000& 0.994& 0.602& 0.417\\
-0.986& -0.964& 0.986& 0.909& 0.694& -0.665& -0.605& 0.973& 0.994& 1.000& 0.683& 0.510\\
-0.722& -0.480& 0.775& 0.919& 0.990& 0.090& 0.146& 0.496& 0.602& 0.683& 1.000& 0.975\\
-0.565& -0.286& 0.623& 0.815& 0.957& 0.300& 0.347& 0.298& 0.417& 0.510& 0.975& 1.000
\end{pmatrix},$
	\end{center}
and the correlation matrix of Solution $C$ is
\begin{center}\label{Scheme IIC}
$\begin{pmatrix}
1.000& 0.929& -0.986& -0.929& -0.662& 0.512& 0.439& -0.924& -0.956& -0.980& -0.785& -0.654\\
0.929& 1.000& -0.886& -0.785& -0.354& 0.788& 0.727& -0.995& -0.988& -0.966& -0.517& -0.340\\
-0.986& -0.886& 1.000& 0.946& 0.724& -0.430& -0.354& 0.885& 0.925& 0.960& 0.834& 0.715\\
-0.929& -0.785& 0.946& 1.000& 0.839& -0.267& -0.199& 0.801& 0.854& 0.911& 0.917& 0.823\\
-0.662& -0.354& 0.724& 0.839& 1.000& 0.285& 0.337& 0.367& 0.462& 0.570& 0.982& 0.997\\
0.512& 0.788& -0.430& -0.267& 0.285& 1.000& 0.941& -0.785& -0.715& -0.623& 0.106& 0.230\\
0.439& 0.727& -0.354& -0.199& 0.337& 0.941& 1.000& -0.713& -0.638& -0.545& 0.171& 0.352\\
-0.924& -0.995& 0.885& 0.801& 0.367& -0.785& -0.713& 1.000& 0.994& 0.973& 0.530& 0.349\\
-0.956& -0.988& 0.925& 0.854& 0.462& -0.715& -0.638& 0.994& 1.000& 0.992& 0.618& 0.447\\
-0.980& -0.966& 0.960& 0.911& 0.570& -0.623& -0.545& 0.973& 0.992& 1.000& 0.711& 0.554\\
-0.785& -0.517& 0.834& 0.917& 0.982& 0.106& 0.171& 0.530& 0.618& 0.711& 1.000& 0.977\\
-0.654& -0.340& 0.715& 0.823& 0.997& 0.230& 0.352& 0.349& 0.447& 0.554& 0.977& 1.000
\end{pmatrix}.$
\end{center}
\end{widetext}
	

\end{document}